\documentclass[a4paper,11pt]{article}

\usepackage{jheppub}

\usepackage[T1]{fontenc}
\usepackage{dsfont}

\title{Yang-Mills mass gap at large-$N$, non-commutative $YM$ theory, topological quantum field theory, and hyperfiniteness}

\author[a,b]{Marco Bochicchio}
\affiliation[a]{INFN sez. Roma 1\\Piazzale A. Moro 2, Roma, I-00185, Italy}
\affiliation[b]{Scuola Normale Superiore (SNS)\\Piazza dei Cavalieri 7, Pisa, I-56100, Italy}
\emailAdd{marco.bochicchio@roma1.infn.it}
\abstract{We review a number of old and new concepts in quantum gauge theories, some of which are well established but not widely appreciated, some are most recent, that may have analogs in gauge formulations of quantum gravity, loop quantum gravity, and their topological versions, and may be of general interest.  Such concepts involve non-commutative gauge theories and their relation to the large-$N$ limit, loop equations and the change to the anti-selfdual variables also known as Nicolai map, topological field theory ($TFT$) and its relation to localization and Morse-Smale-Floer homology, with an emphasis both on the mathematical aspects and the physical meaning. These concepts, assembled in a new way, enter a line of attack to the problem of the mass gap in large-$N$ $SU(N)$ $YM$, that is reviewed as well. \par
Algebraic considerations furnish a measure of the mathematical complexity of a complete solution of large-$N$ $SU(N)$ $YM$: 
In the large-$N$ limit of pure $SU(N)$ $YM$ the ambient algebra of Wilson loops is known to be a type $II_1$ non-hyperfinite factor. 
Nevertheless, for the mass gap problem at the leading $1/N$ order, only the subalgebra of local gauge-invariant single-trace operators matters. The connected two-point correlators
in this subalgebra must be an infinite sum of propagators of free massive fields, since the interaction is subleading in $\frac{1}{N}$, a vast simplification.
It is an open problem, determined by the grow of the degeneracy of the spectrum, whether the aforementioned local subalgebra is in fact hyperfinite. 
Moreover, the sum of free propagators that occurs in the two-point correlators in the aforementioned local subalgebra must be asymptotic for large momentum to the result implied by the asymptotic freedom and the renormalization group: This fundamental constraint fixes asymptotically the residues of the poles of the propagators in terms of the mass spectrum and of the anomalous dimensions of the local operators. \par
For the mass-gap problem, in the search of a hyperfinite subalgebra containing the scalar sector of large-$N$ $YM$, a major role is played by the existence of a $TFT$ underlying the large-$N$ limit of $YM$, with twisted boundary conditions on a torus or, what is the same by Morita duality, on a non-commutative torus. 
The $TFT$ is trivial at the leading large-$N$ order and localized on a set of critical points by means of a quantum version of Morse-Smale-Floer homology, that involves loop equations in the anti-selfdual variables. A hyperfinite sector arises by fluctuations around the trivial $TFT$, in which the joint spectrum of scalar and pseudoscalar glueballs is linear in the square of the masses $m_k^2= k \Lambda_{YM}^2$ with degeneracy $k=1,2, \cdots$, and the two-point correlator satisfies the aforementioned fundamental constraint arising by the asymptotic freedom and the renormalization group.}

\def\beq{\begin{equation}}
\def\eeq{\end{equation}}
\def\bea{\begin{eqnarray}}
\def\eea{\end{eqnarray}}
\def\bq{\begin{quote}}
\def\eq{\end{quote}}

\usepackage{hyperref}

\usepackage{slashed}

\renewcommand{\epsilon}{\varepsilon}
\begin{document}

\maketitle
\flushbottom

\section{Introduction}

The aim of this paper is to review several old and new concepts in quantum gauge theories, some of which are well established but not widely appreciated, some are most recent. \par
The mathematical style of the presentation reflects the nature of the workshop \emph{Mathematical Foundations of Quantum Field Theory}, Jan 16-20 (2012), held at the Simons Center,
but much emphasis is also given to the physical meaning and implications. This paper has been written as a byproduct of the aforementioned workshop, but it has been later expanded to include most recent developments. \par
This short review may be of more general interest, because the old and new techniques described may have analogs, yet to be developed, in formulations of quantum gravity, loop quantum gravity,
and their topological versions (see for example \cite{Vafa}), that involve gauge fields. \par
Despite some of the concepts reviewed have a history that started decades ago, they are assembled in a new way, that is the basis of a line of attack to the problem of the mass gap in large-$N$ $SU(N)$ Yang-Mills ($YM$), and this is the rationale for the style of the exposition, with new asymptotic estimates for glueball propagators entering crucially. From this point of view this paper clarifies in detail which are the real technical difficulties involved in the mass-gap problem, that are rarely, if ever, discussed in the literature. \par

\section{From the loop equation and strings, to the local algebra in large-$N$ $YM$, and hyperfiniteness} \label{s1}

It is known\footnote{See pdf of Detlev Buchholz talk at the Simons Center workshop \emph{Mathematical Foundations of  Quantum Field Theory}, Jan 16-20 (2012), hereby referred to as "this workshop".} that in a $4d$ quantum field theory with a finite number of fields, under mild assumptions on the existence of the $KMS$ states
for any temperature, the von Neumann algebra of the observables is algebraically isomorphic to the unique type $III_1$ hyperfinite factor. \par
We recall that a von Neumann algebra is hyperfinite if it is the weak limit of a sequence of matrix algebras. \par
The situation gets more involved in the large-$N$ limit of any field theory that carries fields in the adjoint representation of $SU(N)$,
in particular in the large-$N$ limit of pure $SU(N)$ $YM$ \cite{Hooft}. \par
The large-$N$ limit of $YM$ is of particular interest, as we will see, because chances are
that we can get non-perturbative information on the real physical spectrum of the particles of the theory and on their interactions, but only in an $\frac{1}{N}$ expansion. \par
In the $YM$ case the large-$N$ limit can be properly defined in terms of the von Neumann algebra generated by Wilson loops $\Psi(x,x;A)$ supported on a loop $L_{xx}$
based at a point $x$:
\bea
\Psi(x,x;A)=P \exp i\int_{L_{xx}} A_{\alpha} dx_{\alpha}
\eea
built by means of the $YM$ connection $A_{\alpha}$.
At leading large-$N$ order Wilson loops satisfy the Makeenko-Migdal loop equation \cite{MM,MM1,Mak1}:
\bea
&&<\frac{1}{2 g^2 N} Tr(\frac{\delta S_{YM}}{\delta A_{\alpha}(x)}\Psi(x,x;A))> = \nonumber \\
&&i \int_{L_{xx}} dy_{\alpha} \delta^{(4)}(x-y) <\frac{1}{N}Tr \Psi(x,y;A)><\frac{1}{N} Tr \Psi(y,x;A)> 
\eea
with fixed 't Hooft coupling \cite{Hooft} $g^2=g^2_{YM} N$, that is a Schwinger-Dyson equation for the Wilson loops in the large-$N$ limit. \par
We can combine the vacuum expectation value (v.e.v) $<...>$  with the normalized matrix trace $\frac{1}{N}Tr$ to define a new normalized trace $TR=<\frac{1}{N}Tr(...)>$ \cite{Si}.
Then the problem is to find an operator solution $A_{\alpha}$ of the Makeenko-Migdal equation uniformly for all loops, with values in a certain operator algebra
with normalizable trace $TR(1)=1$. Such a solution is called the master field \cite{W3}.
Such an algebra is of type $II_{1}$ because of the existence of the normalizable trace, and it is explicitly known. \par
Indeed, the ambient von Neumann algebra of the Makeenko-Migdal loop equation is the Cuntz algebra in its tracial representation
with at least as many self-adjoint generators as the number of components of the gauge connection, i.e. four in $4d$ \cite{Haa1,Haa2,Cv,Douglas,Gross,Douglas1}. \par
The tracial representation of the Cuntz algebra is defined as follows \cite{Douglas,Gross,Douglas1,Voiculescu}:
\bea
a_i  a_j^* &&= \delta_{i j} 1 \nonumber \\
a_i |\Omega> &&=0 \nonumber \\
\sum_i a_i ^* a_i && = 1- |\Omega><\Omega| \nonumber \\
TR(...) && = <\Omega|...|\Omega>
\eea
The construction of the master field in terms of the Cuntz algebra \cite{Douglas,Gross,Douglas1,Voiculescu} involves only four generators since, by a version of the Eguchi-Kawai reduction at large-$N$ \cite{EK,Neu,Twc,Twl1,Twl2,Rt,Mak1,Mak2},
translations can be absorbed by gauge transformations \cite{DN,AG}:
\bea
S A_{\mu} S^{-1} &&=a^*_{\mu}+M_{\mu \nu}a_{\nu}+M_{\mu \nu \rho} a_{\nu} a_{\rho}+... 
\eea
However, the finite number of generators is only seemingly a simplification \cite{MB3}.
In fact, by Voiculescu work \cite{Voiculescu} the von Neumann algebra of the Cuntz algebra with more than one self-adjoint generator in its tracial representation is a type $II_1$ non-hyperfinite  von Neumann algebra, that is algebraically isomorphic to a free-group factor with the same number of generators, that is the main example of the elusive non-hyperfinite type $II_1$ factors.
Therefore, solving the Makeenko-Migdal loop equation is, to use just an euphemism, very difficult \cite{MB3}. \par
We should add that
the von Neumann algebra generated by the actual solution need not to be non-hyperfinite (a string solution \cite{Po1,Po2,Pol} ?), but there is no field-theoretical reason why it should not. \par
So far we described the difficulty of a large-$N$ solution from the abstract algebraic point of view. We now add some physical considerations.
The large-$N$ Makeenko-Migdal loop equation admits a string version \cite{MM,MM1,Mak1} in which the abstract loop equation for the master field with values in the operator algebra of the gauge theory is replaced by a
non-linear functional differential equation for the v.e.v. of Wilson loops $ W(L_{xy})= TR  \Psi(x,y;A)$ (see \cite{Mak1} for the notation):
\bea
\partial^{x}_{\beta} \frac{\delta}{\delta \sigma_{\beta \alpha}} W(L_{xx})=i g^2 \int_{L_{xx}} dx_{\alpha} \delta^{(4)}(x-y) W(L_{xy}) W(L_{xy})
\eea
It is widely believed that the loop equation in this string version admits a string solution, in which the v.e.v. of Wilson loops is realized as a certain path integral of a $2d$ string theory  \cite{Po1,Po2,Pol}
on a world sheet with the topology of a disk whose boundary is the loop. \par
Physically, the hyperfiniteness property can be interpreted as a condition on the number of local degrees of freedom of the theory. 
Indeed, hyperfiniteness is only slightly weaker than existence of the $KMS$ states for any temperature \footnote{See pdf of Detlev Buchholz talk at this workshop.}. \par
For example, the bosonic string does not satisfy the $KMS$ condition for all temperatures
because of the Hagedorn transition \cite{Hagedorn}, i.e. the divergence of the partition function at a certain finite temperature, that is due to the exponential grow of the degeneracy of the spectrum. However, the Hagedorn transition has been related to the tachyon of the bosonic string (see for example \cite{Hagedorn1,Hagedorn2}), in such a way that 
the spectrum of large-$N$ $YM$ may or may not arise by a hyperfinite algebra even in the likely case that a stringy description (obviously non-tachyonic) does exist. \par
This concept of gauge fields/strings duality \cite{Po1} is believed to be an exact equivalence, in the sense that any observable of the gauge theory has an exact correspondence on the string side and vice versa. \par
It is therefore clear that both problems, i.e. solving the Makeenko-Migdal loop equation on the gauge or on the string side, are equally difficult because of the supposed exact equivalence. \par
However, from a practical point of view the Makeenko-Migdal loop equation can be solved on the gauge side order by order in perturbation theory around $g=0$ \cite{Pol}, i.e. at weak coupling, while for the last twenty years it has  been conjectured its solution on the string side 
around $g=\infty$, i.e. at strong coupling, for a certain class of conformal gauge theories with extended supersymmetry ($SUSY$), and for some non-conformal less-supersymmetric deformations \cite{Mal}. \par
This string approach is known under the name of gauge fields/$AdS$ strings correspondence or gauge/gravity duality \cite{Mal}, because in its simplest conformal-supersymmetric incarnation involves a string theory on a $5d$ anti- de Sitter ($AdS$) background, that at strong coupling on the gauge side reduces to the (super)gravity approximation on the string side \cite{Mal}. \par
Nevertheless, both the perturbative approach on the gauge side, and the actual strong-coupling approach on the string side in the present formulation of the $AdS$ strings/gauge fields correspondence, are not helpful for the mass-gap problem \cite{AJ} in large-$N$ $YM$, and
more generally for the spectral problem in any large-$N$ confining asymptotically-free gauge theory with no perturbative mass scale as well, for the reasons explained here below \cite{MB0}. \par
We first recall that the mass-gap problem is the most fundamental problem in pure $YM$ theory, and in its phenomenological application to the theory of strong interactions known as $QCD$ as well. \par
To say it in a nutshell, the mass-gap problem requires to explain why the massless particles that occur in pure $YM$ perturbation theory, the gluons, have never been observed experimentally. Its conjectured solution implies  instead that only massive particles, the glueballs, created from the vacuum by gauge-invariant operators, occur in the theory. \par
The mass-gap problem \cite{AJ} is very difficult \cite{MB0} in pure $YM$, or in any confining asymptotically-free gauge theory with no mass scale in perturbation theory \footnote{These theories include $\mathcal{N}=1$ $SUSY$ $YM$ and $QCD$ with massless quarks. In the last case the theory is believed to have no mass gap since the pion is massless because of the spontaneous breaking of the chiral symmetry, but a mass gap still occurs in the pure-glue sector in the large-$N$ limit.} as well, because the renormalization group ($RG$) together with the asymptotic freedom
($AF$) require that any mass scale of the theory that has a physical meaning, such as the mass gap,
must depend on the canonical coupling constant $g_{YM}$
only through the $RG$-invariant scale $\Lambda_{YM}$, in such a way that in some renormalization scheme, say in the $\overline{MS}$ scheme \cite{schema}:
\bea \label{12}
m_{gap}&=& const  \Lambda_{YM} \nonumber \\
\Lambda_{YM} &=& \Lambda \exp(-\frac{1}{2\beta_0 g_{YM}^2}) (\beta_0 g_{YM}^2)^{-\frac{\beta_1}{2 \beta_0^2}}(1+ \cdots)
\eea
where the $\overline{MS}$ scheme is defined by the relation:
\bea \label{sc}
\log(\frac{\Lambda}{\Lambda_{\overline{MS}}})^2=2 \int_{g_{YM}(\Lambda_{\overline{MS}})}^{g_{YM}(\Lambda)}\frac{dg_{YM}}{\beta(g_{YM})}=\frac{1}{\beta_0 g_{YM}^2(\Lambda)}+\frac{\beta_1}{\beta_0^2}\log g_{YM}^2(\Lambda)+ C +\cdots \nonumber \\
\eea
with $C=\frac{\beta_1}{\beta_0^2} \log\beta_0$, in order to cancel \cite{schema} the term proportional to $\frac{1}{\log^2\frac{\Lambda^2}{\Lambda_{\overline {MS}}^2}}$ in the solution for $g_{YM}$.
Physically, the continuum limit is defined removing the cutoff $\Lambda \rightarrow \infty$ sending at the same time $g_{YM} \rightarrow 0$, in such a way that $\Lambda_{YM}$ is kept constant.
In Eq.(\ref{12}) the result implied by the two-loop beta function is explicitly displayed, while the dots refer to the scheme-dependent higher-loop contributions irrelevant in the ultraviolet. \par
Eq.(\ref{12}) in turn implies that an amazing asymptotic accuracy, as $g_{YM}$ vanishes when the cutoff $\Lambda$ diverges, is needed to solve the mass-gap problem,
and that the mass gap is zero to every order of perturbation theory, since the Taylor expansion of Eq.(\ref{12}) around $g_{YM}=0$ is identically zero. Besides, Eq.(\ref{12}) requires by consistency to sum to all orders the perturbative expansion associated to observables involving any physical mass scale, since perturbative corrections being polynomial in $g_{YM}^2$ are much larger for small $g_{YM}^2$ than the dimensionless function of the coupling in Eq.(\ref{12}). \par
Therefore, a finest asymptotic accuracy of non-perturbative type is needed to get control over the mass gap. \par
This rules out any perturbative method, since the mass gap
is identically zero to every order of perturbation theory. \par
This rules out also any strong coupling method, in particular the strong coupling method involved in the present formulation of the gauge fields/$AdS$ strings duality, because the mass gap has nothing to do with the coupling being large,
since Eq.(\ref{12}) implies that the existence of the mass gap in the continuum limit, i.e. for arbitrarily-large cutoff $\Lambda$, requires an estimate uniformly in a neighborhood of zero coupling asymptotic to Eq.(\ref{12}). \par
In fact, strong coupling methods do not allow us to remove the cutoff,
since if the coupling is very large, $\Lambda_{YM}$ that occurs in Eq.(\ref{12}) is on the order of the cutoff, and then in absence of a uniform estimate that extends to any positive arbitrarily-small neighborhood of zero coupling, the continuum limit cannot be taken, and the proof is lacking that the would-be mass gap survives the continuum limit and it is not an artifact of the finite cutoff introduced in the theory at strong coupling. \par
Another unfit feature of any strong coupling approach is that implicitly assumes
that the strong coupling expansion in a neighborhood of $g_{YM}=\infty$ is connected to the $RG$ flow of the $AF$ theory and that computes meaningful numbers. \par
As a result of the previous arguments we are tempted, or better we are forced, to give up the idea of solving the large-$N$ limit for all the observables of the theory, both on the gauge side and on the string side, in favor of a more limited but more approachable line of attack. \par
Indeed, connected two-point correlators of local single-trace gauge-invariant operators $O^{(s)}(x)$ of spin $s$,
in large-$N$ $YM$ and $QCD$, are in a sense the most simple as possible, a sum of an infinite number of propagators of free fields \cite{M1,M2}
at the next-to-leading $\frac{1}{N}$ order (at the leading order they vanish by standard $\frac{1}{N}$ counting):
\bea \label{KL}
 \int e^{ipx} < O^{(s)}(x) O^{(s)}(0)>_{conn} d^4x 
=\sum_{k} P^{(s)}(\frac{p_{\alpha}}{m_k}) \frac{R^{(s)}_k}{p^2+m_k^2}
\eea
since the interaction is further subleading in the $\frac{1}{N}$ expansion.\par
It is an interesting problem \footnote{This problem has been proposed by the author at this workshop as one of the interesting open problems in quantum field theory.} whether the corresponding local algebra is hyperfinite. In particular, for the problem of the mass gap is most relevant whether the scalar subalgebra is
hyperfinite, since the mass gap is believed to be associated to scalar glueballs on the basis of numerical lattice gauge-theory computations. \par 
It has been known for long that the sum of free fields in Eq.(\ref{KL}) must saturate the logarithms of perturbation theory \cite{M1,M2}, but recently it has been proved an asymptotic structure theorem \cite{MBN}
that determines the residues of the poles in terms of the (asymptotic) spectrum and of the anomalous dimension of the operator $O^{(s)}(x)$.
The asymptotic theorem reads as follows: \par
The connected two-point Euclidean correlator of a local gauge-invariant single-trace operator (and of a fermion bilinear in large-$N$ $QCD$) $\mathcal{O}^{(s)}$ of integer spin $s$ and naive mass dimension $D$ and with anomalous dimension $\gamma_{\mathcal{O}^{(s)}}(g)$,
must factorize asymptotically for large momentum, and at the leading non-trivial order in the large-$N$ limit, over the following poles and residues (after analytic continuation to Minkowski space-time):
\bea \label{at1}
\int \langle \mathcal{O}^{(s)}(x) \mathcal{O}^{(s)}(0) \rangle_{conn}\,e^{-ip\cdot x}d^4x
\sim \sum_{n=1}^{\infty}  P^{(s)} \big(\frac{p_{\alpha}}{m^{(s)}_n}\big) \frac{m^{(s)2D-4}_n Z_n^{(s)2}  \rho_s^{-1}(m^{(s)2}_n)}{p^2+m^{(s)2}_n  } \nonumber \\
\eea
where $ P^{(s)} \big( \frac{p_{\alpha}}{m^{(s)}_n} \big)$ is a dimensionless polynomial in the four momentum $p_{\alpha}$ that projects on the free propagator of spin $s$ and mass $m^{(s)}_n$ and:
\bea \label{g}
\gamma_{\mathcal{O}^{(s)}}(g)= - \frac{\partial \log Z^{(s)}}{\partial \log \mu}=-\gamma_{0} g^2 + O(g^4)
\eea 
with $Z_n^{(s)}$ the associated renormalization factor computed on shell, i.e. for $p^2=m^{(s)2}_n$:
\bea \label{z}
Z_n^{(s)}\equiv Z^{(s)}(m^{ (s)}_n)= \exp{\int_{g (\mu)}^{g (m^{(s)}_n )} \frac{\gamma_{\mathcal{O}^{(s)}} (g)} {\beta(g)}dg}
\eea
The sum in the right-hand side ($RHS$) of Eq.(\ref{at1}) is in fact badly divergent, but the divergence is a contact term, i.e. a polynomial of finite degree in momentum. Thus the infinite sum in the $RHS$ of Eq.(\ref{at1}) makes sense only after subtracting the contact terms (see remark below Eq.(\ref{at2})). Fourier transforming Eq.(\ref{at1}) in the coordinate representation
projects away for $x\neq 0$ the contact terms and avoids convergence problems:
\bea \label{at0}
\langle \mathcal{O}^{(s)}(x) \mathcal{O}^{(s)}(0) \rangle_{conn} 
\sim \sum_{n=1}^{\infty} \frac{1}{(2 \pi)^4} \int  P^{(s)} \big(\frac{p_{\alpha}}{m^{(s)}_n}\big) \frac{m^{(s)2D-4}_n Z_n^{(s)2} \rho_s^{-1}(m^{(s)2}_n)}{p^2+m^{(s)2}_n  } \,e^{ip\cdot x}d^4p \nonumber \\
\eea
The proof of the asymptotic theorem reduces to showing that Eq.(\ref{at1})
matches asymptotically for large momentum, within the universal leading and next-to-leading logarithmic accuracy,
the $RG$-improved perturbative result implied by the Callan-Symanzik equation:
\bea \label{CS}
&& \int \langle \mathcal{O}^{(s)}(x) \mathcal{O}^{(s)}(0) \rangle_{conn}\,e^{-ip\cdot x}d^4x  \nonumber \\
&& \sim P^{(s)}\big(\frac{p_{\alpha}}{p}\big) \, p^{2D-4}    \Biggl[\frac{1}{\beta_0\log (\frac{p^2}{\Lambda^2_{QCD} } )}\biggl(1-\frac{\beta_1}{\beta_0^2}\frac{\log\log (\frac{p^2}{\Lambda^2_{QCD} } ) }{\log (\frac{p^2}{\Lambda^2_{QCD} } )}    + O(\frac{1}{\log (\frac{p^2}{\Lambda^2_{QCD} } )} ) \biggr)\Biggr]^{\frac{\gamma_0}{\beta_0}-1}
\eea
up to contact terms (i.e. distributions supported at coinciding points), and that this matching fixes uniquely the universal asymptotic behavior of the residues in Eq.(\ref{at1}).
More precisely, the asymptotic behavior of the residues is fixed by the asymptotic theorem within the universal, i.e. the scheme-independent, leading and next-to-leading logarithmic accuracy.
This implies that the renormalization factors in the residues are fixed asymptotically for large $n$ to be:
\begin{equation}\label{z1}
Z_n^{(s)2}\sim 
\Biggl[\frac{1}{\beta_0\log \frac{ m^{ (s) 2}_n }{ \Lambda^2_{QCD} }} \biggl(1-\frac{\beta_1}{\beta_0^2}\frac{\log\log \frac{ m^{ (s) 2}_n }{ \Lambda^2_{QCD} }}{\log \frac{ m^{ (s) 2}_n }{ \Lambda^2_{QCD} }}    + O(\frac{1}{\log \frac{ m^{ (s) 2}_n }{ \Lambda^2_{QCD} } } ) \biggr)\Biggr]^{\frac{\gamma_0}{\beta_0}}
\end{equation}
where $\beta_0, \beta_1,\gamma_0$ are the first and second coefficients of the beta function and the first coefficient of the anomalous dimension respectively, and $\Lambda_{QCD}$
is the $QCD$ \footnote{We refer mainly to pure $YM$ or to $QCD$ with massless quarks, but in fact the asymptotic theorem applies to massless $\mathcal{N}$ $=1$ $SUSY$ $QCD$ and more generally to any large-$N$ confining asymptotically-free gauge theory massless in perturbation theory.} $RG$-invariant scale in some scheme. 
Eq.(\ref{at1}) for the propagator can be rewritten equivalently as:
\bea \label{at2}
\int \langle \mathcal{O}^{(s)}(x) \mathcal{O}^{(s)}(0) \rangle_{conn}\,e^{-ip\cdot x}d^4x  
\sim    P^{(s)} \big(\frac{p_{\alpha}}{p} \big)  \, p^{2D-4}   \sum_{n=1}^{\infty} \frac{Z_n^{(s)2}   \rho_s^{-1}(m^{(s)2}_n)  }{p^2+m^{(s)2}_n  }
\eea
up to (divergent) contact terms, where now the sum in the $RHS$ is convergent for $\gamma'=\frac{\gamma_0}{\beta_0} > 1$. Otherwise, it is divergent but the divergence is again a contact term. $P^{(s)} \big(\frac{p_{\alpha}}{p} \big)$ is the projector obtained substituting $-p^2$  to $m_n^2$ in  $P^{(s)} \big(\frac{p_{\alpha}}{m_n} \big)$\footnote{We use Veltman conventions for Euclidean and Minkowskian propagators of spin $s$ \cite{Velt2}.}. \par
An important corollary \cite{MBN} of the asymptotic theorem is that to compute the asymptotic behavior we need not to know explicitly neither the actual spectrum nor the asymptotic spectral distribution, since it cancels by evaluating the sum in Eq.(\ref{at2}) by the integral that occurs as the leading term in  Euler-MacLaurin formula. Hence $RG$-improved perturbation theory does not contain in fact spectral information \cite{MBN}, as perhaps expected. \par
In order to get spectral information it is necessary to abandon asymptotic arguments based only on the Kallen-Lehmann representation and the $RG$ in favor of new field-theoretical methods, whose basic features are described in the next section. \par
But the asymptotic theorem sets the strongest constraints \cite{MBN,SM}
on any solution of the mass-gap problem in the large-$N$ limit, and provides the main tool to falsify any present \cite{MBN,SM} or forthcoming proposed solution. \par
Besides, the fact that at the leading non-trivial order the two-point gauge-invariant local correlators arise by free fields suggests a way-out to the difficulties of a complete large-$N$ solution, 
if a way can be found of getting information only on some scalar correlators, in the likely-hyperfinite local scalar subalgebra at next-to-leading $\frac{1}{N}$ order, avoiding solving for the non-local, non-free, and likely non-hyperfinite algebra of all Wilson loops at leading order. \par 

\section{From hyperfinitess to non-commutative gauge theory and topological field theory} \label{s2}

Therefore, we may wonder as to whether there exist special Wilson loops that generate "small" subalgebras. 
Moreover, if we are interested only in the problem of the mass gap \cite{AJ}, we may limit to correlators that involve only scalar states and possibly to hyperfinite subalgebras \cite{MB0}. \par
At this stage the concept of topological field theory ($TFT$) starts to play a role, because it is naturally associated to such small subalgebras of special Wilson loops, as we will see momentarily. \par
In fact, Witten argued \footnote{See Edward Witten talk at this workshop.} that every gauge theory with a mass gap should contain a possibly trivial $TFT$ in the infrared.\par
Thus we search for a $TFT$ underlying large-$N$ $YM$ whose Wilson loops be trivial at the leading large-$N$ order, but whose correlators for separated loops be non-trivial at next-to-leading order. In our case triviality at the leading large-$N$ order at all scales, and not only in the infrared, is a key feature, because it allows us to bypass the apparent insurmountable difficulty of solving for the algebra of all Wilson loops
at leading order. \par
The triviality of the topological Wilson loops that we look for, i.e. the fact that their v.e.v is $1$, implies the topological property, i.e. the invariance for deformations of the loop, that turns out to be crucial. \par
 Surprisingly, such topological Wilson loops do exist in pure $YM$. To construct them it is necessary a short detour. \par
 It has been known for long, thanks to 't Hooft \cite{HH}, that $YM$ theory can be defined on a torus with twisted, instead of periodic, boundary conditions. 
 Of course, in the thermodynamic limit and at zero temperature, the boundary conditions should not matter, since pure $YM$ is believed to exist in just one phase at zero temperature, the confining asymptotically-free phase characterized by the mass gap. \par
 But $U(N \hat N)$ $YM$ on a torus with twisted boundary conditions is exactly equivalent \cite{Szabo2,AG} by Morita duality to $U(N)$ $YM$ on a non-commutative torus with periodic boundary conditions and non-commutativity $\theta= (L\hat N)^2 \frac{\hat M}{\hat N}$, where $L$ is the side of the commutative torus. \par
 Thus gauge theories on non-commutative space-time, that we refer to as non-commutative gauge theories for brevity, play a role, because the $TFT$ can be naturally defined in the non-commutative setting. \par
Indeed, there exist special Wilson loops, that we call twistor Wilson loops for geometrical reasons explained below \cite{MB0,MB1,Pos1,Pos2}, defined in pure $U(N)$ $YM$ on non-commutative space-time $T^2 \times T^2_{\theta}$ or  $R^2\times R^2_{\theta}$, with complex coordinates $(z, \bar z, \hat u, \hat {\bar u})$, built by a modified non-Hermitian connection $\hat B_{\lambda}$ in the adjoint representation of the Lie algebra:
\bea
\Psi(\hat B_{\lambda};L_{ww})= P \exp i \int_{L_{ww}}(\hat A_z+\lambda \hat D_u) dz+(\hat A_{\bar z}+ \lambda^{-1} \hat D_{\bar u}) d \bar z 
\eea
that define the observables of a trivial $TFT$ underlying the large-$N$ limit of $YM$ \cite{MB0}:
\bea
TR \Psi(\hat B_{\lambda};L_{ww})     & = & TR \Psi(\hat B_1;L_{ww}) \nonumber \\
 \lim_{\theta \rightarrow \infty} TR(\Psi(\hat B_{\lambda};L_{ww})  &=&1 \nonumber \\
\eea
with $\hat D_u=\hat \partial_u+i \hat A_u$ the covariant derivative, and $[\hat u, \hat {\bar u}]=\theta 1, [\partial_{\hat u},\partial_{ \hat {\bar u}}]=\theta^{-1} 1$. \par
Moreover, it is known that  non-commutative $YM$
in the limit of infinite non-commutativity coincides with the large-$N$ limit of commutative $YM$, in such a way that the non-commutative theory realizes in the $\theta \rightarrow \infty$ limit the same master field of the commutative one \cite{Mak1,Mak2,Szabo2,AG}. \par
The $TFT$ is trivial because the generalized trace $TR$ is exactly $1$ for all the topological twistor Wilson loops for $\theta \rightarrow \infty$. 
The trivial topological theory exists at all scales, and not only in the infrared.   \par
The triviality of twistor Wilson loops follows from the fact that in the large-$\theta$ limit they are gauge equivalent to ordinary Wilson loops supported on Lagrangian submanifolds of twistor space of complexified space-time:
\bea
&& P \exp i \int_{L_{ww}}( A_z(z, \bar z, i \lambda z, i \lambda^{-1}\bar z)
+i \lambda  A_u(z, \bar z, i \lambda z, i \lambda^{-1}\bar z)  ) dz \nonumber \\
&&+ ( A_z(z, \bar z, i \lambda z, i \lambda^{-1}\bar z)
+i \lambda^{-1}  A_u(z, \bar z, i \lambda z, i \lambda^{-1}\bar z)  )d \bar z  
\eea
with the parameter $\lambda$ playing the role of the (Lagrangian) fiber of the twistor fibration \cite{MB0}.
In the language of local wedge algebras \footnote{See pdf of Detlev Buchholz talk at this conference.} these loops are supported on (the analytic continuation of) Lagrangian wedges.
Remarkably, the support property implies triviality \cite{MB0}, because of the vanishing of the coefficients of the effective propagators on the support:
\bea
&&\dot z  \dot{\bar z} + i^2 \lambda \dot z  \lambda^{-1} \dot{\bar z} =0 \nonumber \\
&&\dot z \bar z + i^2 \lambda \dot z \lambda^{-1} \bar z=0 \nonumber \\
&& z  \dot{\bar z} + i^2 \lambda z  \lambda^{-1}\dot{\bar z}=0
\eea
The triviality implies a remarkable localization property in function space
of the algebra of twistor Wilson loops \cite{MB1,MB0} in the large-$N$ limit, discussed in the next section.

\section{From the topological field theory to the $YM$ mass gap, via localization by the loop equation in the anti-selfdual variables}

The invariance of twistor Wilson loops for deformations, that arises by their large-$N$ triviality, can be used to localize them on certain critical points of a quantum effective action  \cite{MB0,MB1,Pos1,Pos2,MBN}.\par
Localization in quantum field theory has a long history in supersymmetric gauge theories \cite{DE,A,Bis1,Bis2,W,W1,Szabo}. 
Indeed, the basic idea of $SUSY$ topological field theories \cite{W1} 
is to view certain functional integrals as cohomology classes associated to a nilpotent differential $Q^2=0$ defined by the $BRS$ charge $Q$ obtained by a
twist of the supersymmetry, with the property $\int Q \alpha=0$ for any $\alpha$. 
Thus these physical $SUSY$ theories contain a very special topological subsector defined by closed forms $Q C=Q S_{SUSY}= 0$ modulo total differentials $Q \alpha$, the cohomology class $[C]$ of $C$.
Moreover, the $SUSY$ topological field theory is often solvable, since the cohomology classes $[C]$ are localized on critical points by means of deformations $Q \alpha$ trivial in cohomology:
\begin{equation}
[C]= \int C e^{-S_{SUSY}} =\lim_{t\rightarrow \infty} \int C e^{-S_{SUSY}-t Q\alpha}
\end{equation}
because the saddle-point approximation for large $t$, being $t$ independent for the class $[C]$, is in fact exact \cite{DE,A,Bis1,W}. The aforementioned localization extends to the whole cohomology ring generated by
the closed forms. Hence a small subset of the observables of the theory is localized on critical points. For a review of cohomological localization see \cite{Szabo}. \par
However, in pure $YM$ such a cohomology does not exist, because of the lack of $SUSY$. \par
Thus the new idea, that we review in the following, is to replace cohomology in function space with homology in space-time, and to develop a quantum version of Morse-Smale-Floer theory \cite{MB0}, that associates to non-trivial homology classes critical points. \par
By standard arguments the v.e.v. of (twistor) Wilson loops in the adjoint representation factorize in the large-$N$ limit into the product of the v.e.v in the fundamental and conjugate representation. \par
To twistor Wilson loops in the fundamental representation the following arguments apply. \par
Firstly, the curvature of the twistor connection is a field of anti-selfdual type $F_{z \bar z}(\hat B_{\lambda})= \frac{1}{2} F^-_{01}+\frac{1}{4} \lambda^{-1}(F^-_{02}+i F^-_{03}) - \frac{1}{4} \lambda (F^-_{02}-i F^-_{03})$, with  $F^-_{\alpha \beta}= F^{\alpha \beta}- \,^{*}\!F^{\alpha\beta}$ and $^{*}$ the Hodge dual 
$ ^{*}{F}_{\alpha \beta}=\frac{1}{2}\epsilon_{\alpha\beta\gamma\delta}F^{\gamma\delta} $. \par
Secondly, this suggests that twistor Wilson loops in the $TFT$ may satisfy a simpler-to-solve loop equation, provided the anti-selfdual variables ($ASD$) $F^-_{\alpha \beta}$ are employed as independent integration variables
in the functional integral that defines the $YM$ theory. \par
The change to the $ASD$ variables is a non-$SUSY$ version \cite{MB1,MB0} of the Nicolai map \cite{Nic1,Nic2,V,V1}: It maps the four components of the gauge connection minus the gauge-fixing condition to the three components of the anti-selfdual part of the curvature:
\bea \label{Nic}
Z&=&\int  \exp \big(-\frac{16\pi^2NQ}{2g^2}-\frac{N}{4g^2} \int tr_f (F_{\alpha\beta}^-)^2d^4 x\big)\delta A
\nonumber \\
1&=&\int \delta(F_{\alpha\beta}^--\mu_{\alpha\beta}^-) \delta\mu_{\alpha\beta}^-
\nonumber \\
Z&=&\int  \exp \big(-\frac{16\pi^2NQ}{g^2}-\frac{N}{4g^2}\int tr_f(\mu_{\alpha\beta}^-)^2d^4 x\big)\delta(F_{\alpha\beta}^--\mu_{\alpha\beta}^-) \delta A      \delta \mu_{\alpha\beta}^- 
\eea
Recently, the change to the $ASD$ variables has been proved to define the same one-particle effective action as in the original variables in $YM$, $QCD$ and $\mathcal{N}=1$ $SUSY$ $YM$ in perturbation theory \cite{BP}, thus solving positively a long-standing issue about the actual equivalence of the two formulations. \par 
Thirdly, employing the change to the $ASD$ variables and the choice of the holomorphic gauge $B_{\bar z}=0$ \cite{MB1,MB0}:
\bea
<...>&&= Z^{-1} \int \delta n \delta \bar n \int_{C_{1}} \delta  \mu'  ... \exp(-\frac{N 8 \pi^2 }{g^2} Q-\frac{N 4}{g^2}  \int Tr_f( \mu \bar \mu)  +Tr_f(n + \bar n)^2 d^4x) 
\nonumber\\
&&\delta(-i F_{B} - \mu-\theta^{-1}1) \delta(-i\partial_{A} \bar D- n) \delta(-i\bar \partial_A D- \bar n) \frac{\delta  \mu } { \delta  \mu' }  
\delta A \delta \bar A \delta D \delta \bar D
\eea
a new holomorphic loop equation \cite{MB1,MB0} arises in the $TFT$ as a Schwinger-Dyson equation in the new variables:
\bea
\int  Tr\frac{\delta }{\delta \mu'(z,\bar z)} (e^{-\Gamma}\Psi(B'; L_{zz})) \delta \mu' &=&0
\eea
It reads:
\bea \label{hol0}
&&TR(\Psi(B_{\lambda}; L^{(1)}_{zz})\frac{\delta \Gamma}{\delta \mu_{\lambda}(z, \bar z)}\Psi(B_{\lambda}; L^{(2)}_{zz})) \nonumber \\
&&=\frac{1}{ \pi } \int_{L_{zz}} \frac{ dw}{z-w} TR\Psi(B_{\lambda}; L^{(1)}_{zw}) TR\Psi(B_{\lambda}; L^{(2)}_{wz})
 \eea
for a loop with the shape of the symbol $\infty$ and two petals $L^{(1)}$ and $L^{(2)}$. For real points \cite{MB0}, or by analytic continuation to Minkowski space-time \cite{MB1}, it can be regularized in a gauge-invariant way:
\bea
&& TR (\Psi(B_{\lambda}; L^{(1)}_{z_+ z_+})\frac{\delta \Gamma}{\delta \mu_{\lambda}(z_+, z_+)}\Psi(B_{\lambda}; L^{(2)}_{z_+ z_+}))  = \nonumber \\
&& \frac{1}{ \pi} \int_{L_{z_+z_+}} (P\frac{dw_+}{z_+ -w_+}+i dw_+\pi \delta(z_+ -w_+)) \nonumber \\
&&TR\Psi(B; L_{z_+w_+}) TR\Psi(B; L_{w_+z_+})
\eea
with the principal part not contributing, being supported on open (twistor) Wilson loops, whose v.e.v. vanishes by gauge invariance \cite{MB0,MB1}. \par
Finally, deforming the loop to a cusped loop with zero cusp angle at the non-trivial self-intersection, 
it follows that the right-hand side of the holomorphic loop equation is exactly zero:
\bea
\int dw_+(s)
\delta(z_+(s_{cusp}) -w_+(s))&=&\frac{1}{2}\frac{\dot w_+(s^+_{cusp})}{ |\dot w_+(s^+_{cusp})|}+ \frac{1}{2}
\frac{\dot w_+(s^-_{cusp})}{|\dot w_+(s^-_{cusp})|} =0
\eea
because of the opposite orientation of the arcs asymptotic to the cusps \cite{MB0,MB1}. \par
This is the localization property, that expresses the fact that the matrix elements of the equation of motion of $\Gamma$ vanish,
when restricted to the subalgebra of twistor Wilson loops:
\bea
TR (\Psi(B; L^{(1)}_{z_+z_+})\frac{\delta \Gamma}{\delta \mu(z_+, z_+)}\Psi(B; L^{(2)}_{z_+z_+}))= 0
\eea
There is a homological interpretation of the localization, such that the holomorphic loop equation is localized by the addition to the loop of vanishing boundaries,
 i.e. backtracking arcs ending with cusps \cite{MB1,MB0, Pos1}, in the dual way to the localization of cohomology classes, that involves deforming by coboundaries \cite{DE,A,Bis1,Bis2,W,W1,Szabo}. \par
 Moreover, globally the existence of the critical points can be interpreted \cite{MB0} as a version of Morse-Smale-Floer homology, provided the cusps occur as double points on a punctured sphere
 with pairwise-identified punctures, and the loops occur as non-trivial homology cycles asymptotic to the cusps on the punctured sphere \cite{MB0}. \par
Locally, on a punctured disk of the sphere, for a lattice of punctures and a dense set in function space \cite{MB0}, by the resolution of identity in Eq.(\ref{Nic}) it holds:
 \bea \label{000}
F_{\alpha \beta}^- = 
\sum_p \mu^{-}_{\alpha \beta}(p)  \delta^{(2)} (z-z_p)
\eea
If the gauge group is unbroken at the critical points, the twistor Wilson loops in the fundamental representation must have $Z_N$ holonomy around the singular points of the lattice divisor.
Thus the localization in the $TFT$ realizes a version of 't Hooft electric/magnetic duality, in which the magnetic charge condenses and $YM$ theory confines the electric charge \cite{MB0,MB1,Pos1,Pos2}. \par
The point-like singularities on the $2d$ surface of the non-commutative gauge theory lift to surface-like singularities \cite{MB0,MB1,Pos1,Pos2} known as surface operators in the corresponding Morita-equivalent commutative gauge theory.
Thus the algebra of observables of the $TFT$ can be realized explicitly as the closure of a dense set in function space,
that involves a lattice of surface operators supported on Lagrangian submanifolds \cite{MB1,MB0}.\par
Mathematically, these are local systems, i.e. representations of the fundamental group of the punctured sphere, associated to the $TFT$, obtained interpreting \cite{MB0} the Nicolai map restricted to the lattice divisor Eq.(\ref{ASD}) as hyper-Kahler reduction \cite{H,HKL}. \par
On the lattice hyper-Kahler quotient the physical fluctuations around the topological theory are locally Abelian in function space, all the other
non-Abelian degrees of freedom being zero modes of the Jacobian of the Nicolai map associated to the moduli of the local system \cite{MB0}. \par
The $TFT$ is locally Abelian \cite{MB0} because of the automatic commutativity, for solutions of Eq.(\ref{000}) \cite{S5,Moc,W2,S6}, of the triple $\mu^{-}_{\alpha \beta}(p)$ at each lattice point $p$, the new integration variables in force of the change to the $ASD$ variables, due to the singular nature of Hitchin equations Eq.(\ref{000}) \cite{H2,S1,S2}. \par
This is the fundamental result \cite{MB0}, that allows us to compute reliably the physical fluctuations around the $TFT$ in the large-$N$ limit, because the number of local fluctuating fields in the $TFT$
is on the order \cite{MB0} of $N$ instead of $N^2$. \par
As an aside, on the lattice hyper-Kahler quotient the critical points of the effective action occur as fixed points \cite{Pos1}
for the action on the $ASD$ variables in function space of the semigroup that rescales the fiber of the Lagrangian twistor fibration,
because of the $\lambda$-independence of the v.e.v. of twistor Wilson loops:
\bea
 TR \Psi( B_{\lambda};L_{ww})  & = & TR \Psi( B_1;L_{ww}) 
\eea
Thus the topological sector is localized at the leading large-$N$ order on the critical points of the effective action \cite{MB0}:
\bea \label{hol}
\Gamma&=&\frac{4N \hat N}{g_W^2} \int d^2u d^2z   \rho^2   tr_N    Tr_{\hat N } (\mu\bar \mu+ \nu^2)+\int d^2u d^2z  \rho^2 (\log \Delta(\mu) 
- \log|\Delta(\nu+ \mu-\bar \mu)|^2 )\nonumber \\
&-& \log Det^{-1/2}(-\Delta_{A} \delta_{\alpha \beta}-i ad \mu_{\alpha \beta}^- ) Det(-\Delta_{A})  -  \int d^2u d^2z   \rho^2 n_b [\mu']  \log \Lambda- \int d^2u \rho \log Det(\omega')^{\frac{1}{2}}  \nonumber \\
&+& \text{complex conjugate}
\eea
where $\nu=n+\bar n$, $\rho$ is the density of surface operators, and $\Delta(\mu)$ the Vandermonde determinant of the eigenvalues of $\mu$ \cite{MB0}. \par
$\Gamma$ contains the interesting information of the localization,
since it is naturally defined on a physical wedge, rather than on the topological one. \par
Indeed, though the topological theory is trivial at large-$N$, the effective action $\Gamma$ admits at subleading $1/N$ order non-trivial fluctuations around the critical points,
supported on a "transverse" Lagrangian wedge, that is a physical wedge:
\bea
(z, \bar z, i \lambda  z, i \lambda^{-1} \bar z)
 \rightarrow (z_+, z_-, -\lambda  z_+, -\lambda^{-1} z_-)
\eea
i.e. a certain Lagrangian wedge, that is defined through the analytic continuation, as operators, of the topological twistor Wilson loops to the physical twistor Wilson loops:
\bea
\Psi(\hat B_{\lambda};L_{ww}) \rightarrow  P \exp i \int_{L_{ww}}(\hat A_{z_+} + i \lambda \hat D_{u_+}) dz_+ + (\hat A_{ z_-}+i\lambda^{-1} \hat D_{ u_-}) dz_- 
\eea
The physical consequences of the localization are as follows. \par
A large-$N$ beta function of Novikov-Shifman-Vainshtein-Zakharov type \cite{NSVZ} at the leading  $1/N$ order follows \cite{MB1,MB0}:
\begin{align} \label{alfa}
&\frac{\partial g}{\partial \log \Lambda}
=\frac{-\beta_0 g^3+\frac{1}{(4\pi)^2}g^3 \frac{\partial \log Z}{\partial \log \Lambda}}{1-\frac{4}{(4\pi)^2}g^2}=-\beta_0 g^3-\beta_1 g^5+\cdots \\
&\frac{\partial g_W}{\partial \log \Lambda}=-\beta_0 g_W^3 \\ \label{beta0}
&\frac{\partial\log Z}{\partial\log \Lambda} =\frac{2\gamma_0 g_W^2}{1+c' g_W^2}=2\gamma_0 g^2 +\cdots  
\end{align}
with $\gamma_0=\frac{1}{(4\pi)^2}\frac{5}{3}$ and $c'$ a scheme-dependent constant. It is easy to check that this beta function reproduces \cite{MB1,MB0} the correct universal one- and two-loop coefficients of the perturbative $YM$ beta function $\beta_0=\frac{1}{(4\pi)^2}\frac{11}{3}$ and  $\beta_1=\frac{1}{(4\pi)^4}\frac{34}{3}$, by noticing that for small $g_W^2$,$\, g_W^2 \sim g^2 $ within the leading logarithmic accuracy, and expanding in powers of $g^2$. \par
At the next to leading $1/N$ order, the mass gap, the glueball spectrum and the hyperfiniteness in the joint scalar and pseudoscalar sector follow as well, for fluctuations
supported on the transverse Lagrangian wedge, from the correlator of surface operators \cite{MB0,MBN}:
\bea \label{ASD}
\int \langle TrF^{-2}(x) TrF^{-2}(0) \rangle_{conn}\,e^{-ip\cdot x}d^4x  
=    \, \frac{2}{\pi^2} p^{4}   \sum_{k=1}^{\infty} \frac{g^4(k \Lambda^2_{YM})   \Lambda^2_{YM} }{p^2+k \Lambda^2_{YM}  }
\eea
This correlator couples to scalar and pseudoscalar glueballs \cite{MB0,MBN}. In the $TFT$ the joint spectrum of scalar and pseudoscalar glueballs in the large-$N$ limit is exactly linear in the masses squared $m_k^2=k \Lambda^2_{YM}$ with degeneracy \cite{MB0} $k$, thus implying the hyperfiniteness property in the $TFT$, since the free partition function in this sector of $YM$ exists for every temperature because of the moderate (polynomial) grow of the degeneracy of the spectrum. \par
The strongest self-consistent check of Eq.(\ref{ASD}) is that satisfies \cite{MBN} the asymptotic theorem. Indeed, the asymptotic theorem implies \cite{MBN, SM}:
\bea 
\int \langle TrF^{-2}(x) TrF^{-2}(0) \rangle_{conn}\,e^{-ip\cdot x}d^4x  
\sim    \, p^{4}   \sum_{n=1}^{\infty} \frac{g^4(m_n^2)   \rho_0^{-1}(m^{2}_n)  }{p^2+m^{2}_n  }
\eea
since $\gamma_0=2\beta_0$ \cite{SM} for the operator $TrF^{-2}$. \par
Morevorer, in order to confirm that strong coupling methods are not helpful for the mass-gap problem in $YM$,
we observe that the residues of the poles of the scalar and pseudoscalar glueball propagators (deprived of the dimensionful factors of $p^4$, and of $\rho_0^{-1}$, that may be assumed to be constant asymptotically because of the (expected) asymptotic linearity of the spectrum of masses squared), vanish as $g^4(m^2_n)$ for large $n$, as required by the asymptotic freedom, while the scale $m_n^2$ of the corresponding poles diverges, in complete disagreement with the idea that the non-vanishing masses $m_n$ occur because the theory becomes strongly coupled: 
The more the theory becomes weakly coupled in the ultraviolet, the more massive poles occur, since they must be infinite in number to match at the leading large-$N$ order the $RG$-improved perturbative behavior, with vanishingly-small dimensionless residues because of the weak coupling. \par
Presently, no other proposal for the scalar or pseudoscalar correlators satisfies the asymptotic theorem \cite{MBN, SM}. 

\section{Acknowledgments}
The hyperfiniteness problem has been suggested by the author as one of several problems arising as a byproduct of the Simons Center
workshop: \emph{Mathematical Foundations of Quantum Field Theory}, Jan 16-20 (2012), but this paper accounts  for most recent developments
occurred after the workshop.\par
We would like to thank Alexander Migdal and Alexander Polyakov for a long fascinating discussion on loop equations in Princeton during the workshop.
In particular one question that arose in that discussion helped us to focus on the local Abelianization of surface operators in the topological field theory,
that is the mathematical and physical justification \cite{MB0} of the results reported in \cite{MBN}.\par 
We would like to thank the organizers and the participants for the invitation to participate,
the stimulating atmosphere and the financial support.
In particular we would like to thank Detlev Buchholz, Mike Douglas, Mark Harder, Arthur Jaffe, Tom Spencer, for very interesting
discussions on quantum field theory and Yang-Mills.

\end{document}